\documentclass[12pt]{article}

\advance\topmargin by -1.6cm

\newenvironment{eq}[1]
{\[\begin{array}{#1}}{\end{array}\]}

\let\rvec=\vec        
\let\Sup=\sup        

 \def\({\Bigl(}
\def\){\Bigr)}   
 \def\|{\Big|}
\def\then{\Rightarrow}  
 \def\o{\circ}
\def\m{\bullet}    
\def\x{\times}
   
\def\ox{\otimes}

\def\pl{{~\oplus~}}

\def\SUM{\displaystyle \sum}

\def\mid{\big\bracevert}

\def\sub{\subseteq}
\def\subnoteq{\subset}
\def\sup{\supseteq}
\def\supnoteq{\supset}
\def\and{\wedge}

\def\rin{{\,\in\kern-.42em\in}}
 
 \def\diag{{\,{\rm diag}\,}}

\def\spec{\,{\rm spec}\,}

\def\det{\,{\rm det }\,}
\def\id{\,{\rm id}}

\def\sx{~\rvec\x~\!}

\def\A{{\,{\rm A\kern-.55emA}}}
\def\B{{\,{\rm I\kern-.2emB}}}
\def\C{{\,{\rm I\kern-.55emC}}}
\def\E{{\,{\rm I\kern-.2emE}}}
\def\G{{\,{\rm I\kern-.55emG}}}
\def\H{{{\rm I\kern-.2emH}}}
\def\I{{\,{\rm I\kern-.2emI}}}
\def\K{{\,{\rm I\kern-.2emK}}}
\def\L{{\,{\rm I\kern-.2emL}}}
\def\M{{\,{\rm I\kern-.16emM}}}
\def\N{{\,{\rm I\kern-.16emN}}}
\def\Q{{\,{\rm I\kern-.5emQ}}}
\def\R{{{\rm I\kern-.2emR}}}
\def\S{{\,{\rm I\kern-.42emS}}}
\def\T{{\,{\rm I\kern-.37emT}}}
\def\UU{{\,{\rm I\kern-.51emU}}}
\def\Z{{\,{\rm Z\kern-.32emZ}}}

\def\p{\partial}

\def\al{\alpha}   \def\ga{\gamma}
\def\de{\delta}  \def\ep{\epsilon}  
\def\th{\theta}   \def\vth{\vartheta} 
      
   \def\om{\omega} \def\Om{\Omega}
\def\phi{\varphi} 
 \def\Ga{\Gamma}  
    \def\La{\Lambda}

\def\vec#1{\underline{\bf vec}_{#1}}

\def\GL{{\bf GL}}  
\def\SL{{\bf SL}}
\def\U{{\bf U}} 
 
\def\O{{\bf O}}   
\def\SU{{\bf SU}} 
 
\def\SO{{\bf SO}}

\def\d#1{{\check{#1}}}
\def\angle#1{\langle#1\rangle}

\def\rstate#1{|#1\rangle}
\def\lstate#1{\langle#1|}

\def\ty#1{{\tt #1}}
\def\ro#1{{\rm #1}}
\def\bl#1{{\bf {#1}}}
\def\cl#1{{\cal #1}}

\def\ol#1{\overline{#1}}

\def\dprod#1#2{\langle#1,#2\rangle}
\def\sprod#1#2{\langle#1|#2\rangle}
\def\rsprod#1#2{\lbrace#1|#2\rbrace}

\def\com#1#2{\lbrack#1,#2\rbrack}

\def\acom#1#2{\{#1,#2\}}

\def\map{\longrightarrow}

\def\lrmap{\leftrightarrow}

\def\mape{\longmapsto}

\begin{document}

\begin{titlepage} 
$~$
\vskip5mm
\hfill MPP-2005-2
\vskip25mm

\centerline{\bf THE HILBERT SPACES}
\vskip5mm
\centerline{\bf  FOR STABLE AND UNSTABLE PARTICLES}
\vskip15mm
\centerline{
Heinrich Saller\footnote{\scriptsize hns@mppmu.mpg.de} }
\centerline{Max-Planck-Institut f\"ur Physik}
\centerline{Werner-Heisenberg-Institut}
\centerline{M\"unchen, Germany}

\vskip25mm

\centerline{\bf Abstract}
\vskip5mm

The Hilbert spaces for stable scattering states and particles are determined 
by the re\-pre\-sen\-ta\-tions of the 
characterizing Euclidean and Poincar\'e group and given,  respectively,  by the square
integrable functions on the momentum 2-spheres for a fixed absolute value
of momentum  and  on the energy-momentum
3-hyperboloids for a particle mass. 
The Hilbert spaces for the corresponding unstable
states and particles are not characterized by square integrable functions.
Their scalar products are defined by positive type functions for the cyclic
re\-pre\-sen\-ta\-tions of the time, space and spacetime translations involved.
Those cyclic, but reducible  translation re\-pre\-sen\-ta\-tions are irreducible as re\-pre\-sen\-ta\-tions of
the corresponding affine operation groups which involve 
also the time, space and spacetime reflection
group, characteristic for unstable structures.

\end{titlepage}

\vskip15mm
{\small\tableofcontents}
\newpage
\setcounter{page}{3}

\section{Introductory Remarks}

In quantum theory, the operational structure of physics, e.g. 
the action of time and space
translations, of rotations or of electrodynamical phase transformations,
is experimentally observed
in terms of   transition amplitudes and probabilities. 
The related complex and positive numbers 
are formulated with the scalar product of  Hilbert spaces
acted upon with the operations groups,
in general real Lie groups.

Each real Lie group determines `its' characteristic complex Hilbert spaces.
E.g., the real $(1+2s)$-dimensional 
Heisenberg group $\bl H(s)$ with $s$ position momentum pairs in
the characterizing Lie algebra bracket $[\bl x_a,\bl p_b]=\de_{ab}\bl I$
determines - with the Stone-von Neumann theorem\cite{NEUM,FOL,S041}  - 
 the square integrable function classes $L^2(\R^s)$
  as the
 Hilbert spaces for its faithful re\-pre\-sen\-ta\-tions
 with characterizing invariant $\hbar\ne0$.
 The $\bl H(s)$-Hilbert spaces are not universal.
E.g. the irreducible Hilbert spaces for nonrelativistic scattering
are proper subspaces of $L^2(\R^3)$. 
In important cases,  there are irreducible Hilbert spaces for
an operation group which are not
formalizable with square integrable functions.  
The Hilbert spaces for unstable structures are examples.

In general, the irreducible sets where a group $G$ acts upon
are $G$-orbits, isomorphic to subgroup classes $G/H$ with the realization
characterizing fixgroup ('little group') $H\sub G$. All elements of an orbit
are cyclic $G\m x\cong G/H$.
Representations act - by definition - on  vector spaces. A nondecomposable 
re\-pre\-sen\-ta\-tion space is the linear span of the orbit 
of a cyclic vector - with the additional closure 
in the case of  topological vector spaces.  
In the following without mentioning the contrary, all re\-pre\-sen\-ta\-tions
considered are in definite unitary groups, also called
Hilbert re\-pre\-sen\-ta\-tions.

Some basic re\-pre\-sen\-ta\-tion theoretical concepts:
Re\-pre\-sen\-ta\-tions of Lie groups are direct sums of cyclic 
re\-pre\-sen\-ta\-tions, i.e. of 
re\-pre\-sen\-ta\-tions with a cyclic vector,  and direct integrals of
irreducible re\-pre\-sen\-ta\-tions\cite{PWEYL,FOL,KIR}. For compact Lie groups, 
cyclic re\-pre\-sen\-ta\-tions are irreducible
and finite dimensional  - not necessarily for noncompact Lie groups
where faithful re\-pre\-sen\-ta\-tions have to be infinite dimensional.

\section{The Hilbert Spaces for Stable Particles}

According to Wigner\cite{WIG}, a  particle is characterized by an  irreducible unitary 
re\-pre\-sen\-ta\-tion of the Poincar\'e group $\SO_0(1,3)\sx\R^4$
as semidirect product $G=H\sx N$
(subgroup $H$ acting upon normal subgroup $N$)
  of the homogeneous orthochronous Lorentz group 
with  the spacetime translations.
The Poincar\'e group  acts upon an infinite dimensional 
Hilbert space with the invariant particle properties spin and mass and the
eigenvalues 3rd spin component and energy-momenta for the eigenvectors.

The Poincar\'e group $\SL(\C^2)\sx\R^4$
for special relativity, with covering group $\SL(\C^2)$ of the homogeneous group
$\SO_0(1,3)\cong \SL(\C^2)/\{\pm\bl1_2\}$,
contains - according to each decomposition into  time
translations  and position translations  -
 direct products $\x$ of time translations 
and  Euclidean groups with rotations
$\SO(3)\cong \SU(2)/\{\pm\bl1_2\}$
acting upon  position  translations
\begin{eq}{l}
\SO_0(1,3)\sx\R^4\supnoteq \R\x [\SO(3)\sx\R^3]
\end{eq}Time and position translation re\-pre\-sen\-ta\-tions, to be given first, are Lorentz
compatibly composed to Poincar\'e group re\-pre\-sen\-ta\-tions.

\subsection 
[- for Energy Eigenstates with Time Translations]
{- for Energy Eigenstates with Time Translations}

With Schur\cite{SCHUR}, the  irreducible   
re\-pre\-sen\-ta\-tions of the abelian time translation group 
are complex $1$-dimensional. They are characterized by an invariant energy $E$
\begin{eq}{l}
\R\ni t\mape e^{iEt}\in\U(1),~~E\in\R
\end{eq}The eigenvector $\rstate E$ for a stable state 
with energy $E$ spans the 1-dimensional
re\-pre\-sen\-ta\-tion space $\C\rstate E$. It
is acted upon with the time translations  which describes its time
development and gives the time orbit in the re\-pre\-sen\-ta\-tion space  - a circle
in  the complex plane 
\begin{eq}{l}
\R\ni t\mape \rstate{E,t}=e^{iEt}\rstate E\in \C\rstate E
,~~\rstate{E,0}=\rstate E
\end{eq}The translation dependence, here  $t$, is included in the ket
and omitted for trivial translations, here for $t=0$.

The time translation representing unitary group $\U(1)$
defines the scalar product of the re\-pre\-sen\-ta\-tion space.
The basic
vector can be normalized (probability 1)
\begin{eq}{l}
\sprod EE=1
\end{eq}The time dependence of the scalar product
(transition or probability amplitude) reproduces  the time 
re\-pre\-sen\-ta\-tion matrix element above
\begin{eq}{l}
\sprod{E,t_2}{E,t_1}=e^{iE(t_1-t_2)}\sprod EE=e^{iEt}\hbox{ with }t=t_1-t_2 
\end{eq}

Irreducible time translation re\-pre\-sen\-ta\-tions with different characterizing
invariant energies  are Schur-orthogonal to each other 
 as seen by integration with Haar measure $dt$ over
the parameters of the translation group
\begin{eq}{l}
\{D^E\mid E\in\R\}\hbox{ with } D^E(t)=e^{iEt}\cr

\rsprod{D^{E_2}}{D^{E_1}}=
\int dt~e^{-iE_2 t}e^{iE_1 t}=\de({E_1-E_2\over 2\pi}) 
\end{eq}

In general, the Schur scalar product\cite{SCHUR,FOL,HELG2,KNAPP}  for 
the irreducible  re\-pre\-sen\-ta\-tion matrix elements
(also called re\-pre\-sen\-ta\-tion coefficients),
 denoted here with the 
`braced bra and ket' $\rsprod{\dots}{\dots}$,  has to be distinguished from
the scalar product for the vectors of an irreducible re\-pre\-sen\-ta\-tion space,
denoted here with the `usual bra and ket' $\sprod{\dots}{\dots}$.

An easily accessible example for this distinction is the compact rotation group
$\SU(2)$ with the  scalar products for the  finite dimensional 
irreducible representation spaces and the Schur scalar product for all $\SU(2)$-functions
$L^2(\SU(2))$: In the Euler parametrized  representations,
e.g. the adjoint 3-dimensional one
\begin{eq}{rl}
\SU(2)\ni u(\chi,\phi,\th)&\mape [2J](\chi,\phi,\th)\in\U(1+2J)\cr
\hbox{e.g. } 
[2]_n^m(\chi,\phi,\th)&={\scriptsize\pmatrix{
e^{i\chi}e^{i\phi}\cos^2{\th\over2}&
ie^{i\phi}{\sin \th\over\sqrt2}&
-e^{-i\chi}e^{i\phi}\sin^2{\th\over2}\cr
ie^{i\chi}{\sin\th\over\sqrt2}&
\cos \th&
ie^{-i\chi}{\sin\th\over\sqrt2}\cr
-e^{i\chi}e^{-i\phi}\sin^2{\th\over2}&
ie^{-i\phi}{\sin \th\over\sqrt2}&
e^{-i\chi}e^{-i\phi}\cos^2{\th\over2}\cr}}
\end{eq}the $(1+2J)$ columns constitute,  for each group element
$u(\chi,\phi,\th)$, 
  a  spherical basis. These $(1+2J)$ vectors can be chosen unitary  
with respect to the representation space scalar product
\begin{eq}{l}
\sprod {J;n'}{J,n}={\SUM_m}
\ol{[J]_{n'}^m(\chi,\phi,\th)}~
{[J]_n^m(\chi,\phi,\th)}=\de_{nn'}
\end{eq}without any integration. Integrating over the group, all  matrix elements
are Schur-orthonormalized
\begin{eq}{rl}
\rsprod{[J']_{n'}^{m'}}{[J']_{n'}^{m'}}
=&\int_{-2\pi}^{2\pi}{d\chi\over 4\pi}
\int_{0}^{2\pi}{d\phi\over 2\pi}
\int_{-1}^{1}{d\cos\th\over 2}
\ol{[J']_{n'}^{m'}(\chi,\phi,\th)}~
{[J]_n^m(\chi,\phi,\th)}\cr
=&{1\over 1+2J}\de_{nn'}\de^{mm'}\cr
\hbox{e.g. }&
\int_{-2\pi}^{2\pi}{d\chi\over 4\pi}
\int_{0}^{2\pi}{d\phi\over 2\pi}
\int_{-1}^{1}{d\cos\th\over 2}|e^{i\chi}e^{i\phi}\cos^2{\th\over2}|={1\over3}
\end{eq}
\subsection
[- for Scattering States with Position Translations]
{- for Scattering States with  Position Translations}

The irreducible  re\-pre\-sen\-ta\-tions of the 
noncompact and nonabelian  Euclidean group
 for position space, a semidirect product of rotations acting upon translations
\begin{eq}{l}
\SU(2)\sx\R^3,~~(O,\rvec x)\in\SO(3)\sx\R^3\cr
(O_1,\rvec x_1)\o (O_2,\rvec x_2)=(O_1O_2,\rvec x_1+O_1\rvec x_2)\cr
\SO(3)\sx\R^3/\SO(3)\cong\R^3\hbox{ (no group isomorphism)}
\end{eq}are more complicated than the 1-dimensional ones 
for the time translation group $\R$.

With Schur again, the irreducible re\-pre\-sen\-ta\-tions of translations 
in any dimension $n=1,2,\dots$
are 1-dimensional with the (energy-)momenta as invariant eigenvalues
(dual group, characters)
\begin{eq}{l}
\R^n\ni x\mape e^{ipx}\in\U(1),~~p\in\R^n
\end{eq}There is a good method to give the   re\-pre\-sen\-ta\-tion of affine groups:
With Wigner and Mackey\cite{WIG,MACK1,MACKP, FOL}, the affine group
re\-pre\-sen\-ta\-tions can be induced from   re\-pre\-sen\-ta\-tions of 
fixgroups (Wigner's  little
groups) which leave  (energy-)momenta invariant.

For position translations $\rvec x\in \R^3$ with  nontrivial momentum eigenvalues 
\begin{eq}{l}
\rvec p\in\R^3\hbox{ with }\rvec p^2=P^2>0
\end{eq}the fixgroup consists of  the axial rotations $\SO(2)$ around 
the momentum direction ${\rvec p\over|\rvec p|}$.
The    $\SO(2)$-re\-pre\-sen\-ta\-tions together with
the   position translation re\-pre\-sen\-ta\-tions induce 
all   re\-pre\-sen\-ta\-tions 
of the Euclidean group $\SU(2)\sx\R^3$. They
 collect all translation re\-pre\-sen\-ta\-tions 
 $\rvec x\mape e^{-i\rvec p\rvec x}$ for  momenta $\rvec p$ on the momentum
sphere $\Om^2\cong\SO(3)/\SO(2)$ with invariant radius $P$. E.g.,
 the scalar  matrix elements
arise with the trivial $\SO(2)$-re\-pre\-sen\-ta\-tion 
\begin{eq}{rl}
\R^3\ni \rvec x\mape& 
\int{d^3 p\over2\pi P}\de(\rvec p^2-P^2)e^{-i\rvec
p\rvec x}={\sin Pr\over Pr}=j_0(Pr)\cr
\hbox{for}& P>0\hbox{ with }r=\sqrt{\rvec x^2}
\hbox{, supported by }\rvec p^2=P^2
\end{eq}They are normalized for the
neutral position translation $\rvec x=0$
(`here'). The value of the spherical Bessel function $j_0(Pr)$ 
is a matrix element
 of an infinite
dimensional re\-pre\-sen\-ta\-tion. It  has to  be seen in analogy  to $e^{iEt}$
as matrix element of a 1-dimensional
time translation  re\-pre\-sen\-ta\-tion (more below). 

As a translation  $\R^3$-re\-pre\-sen\-ta\-tion, the matrix element 
$j_0(Pr)$ belongs to a cyclic re\-pre\-sen\-ta\-tion.
It is not decomposable into a direct sum of 
irreducible $\R^3$-re\-pre\-sen\-ta\-tions, but into 
a direct integral.
As an $\SO(3)\sx\R^3$-re\-pre\-sen\-ta\-tion
it is irreducible as seen from the nontrivial rotation behavior 
for the irreducible components $e^{-i\rvec p\rvec x}
\mape e^{-i\rvec p (O.\rvec x)}$. With the homogeneous group 
acting upon the translations, all momenta on a 2-sphere have
to be included as eigenvalues of the translation
re\-pre\-sen\-ta\-tion $\{e^{i(O.\rvec p)\rvec x}\mid O\in\SO(3)\}$.

Starting from flat position space $\R$ with trivial rotations $\SO(1)=\{1\}$,
the  $\SO(3)$-rotation degrees of freedom 
use - via the $\Om^2$-derivative ${\p \over\p {r^2\over4\pi}}$ -  the 2-sphere spread 
of a 1-dimensional translation re\-pre\-sen\-ta\-tion matrix element 
\begin{eq}{l}
\int d^3 p~\de(\rvec p^2-1)e^{-i\rvec
p\rvec x}=
-{\p \over\p {r^2\over4\pi}} \int d p~\de(p^2-1)e^{-ipr}
=-{\p \over\p {r^2\over4\pi}} \cos r=2\pi j_0(r)
\end{eq}Matrix elements for a nontrivial re\-pre\-sen\-ta\-tion of the 
homogeneous rotation
group $\SO(3)$ are obtained by position derivations 
${\p\over\p\rvec x}={\rvec x\over r}{\p\over\p r}=2\rvec x{\p\over\p r^2}$, e.g.
with angular momentum $L=1$ involving the spherical harmonics
$\ro Y_{1,2,3}^1(\phi,\th)\sim{\rvec x\over r}$
multiplied with the matching spherical Bessel function $j_1$
\begin{eq}{l}
\R^3\ni \rvec x\mape 
\int{d^3 p\over2\pi P}~i{\rvec p\over P}\de(\rvec p^2-P^2)
e^{-i\rvec p\rvec x}={\rvec x\over r}j_1(Pr),~~j_1(r)={\sin r-r\cos r\over r^2}
\end{eq}The matching products
$\rvec x\mape \ro Y^L_m(\phi,\th)j_L(Pr)$ of spherical harmonics and spherical Bessel  functions
are familiar e.g. from the planar wave decomposition
into matrix elements of irreducible $\SO(3)\sx\R^3$-re\-pre\-sen\-ta\-tions
\begin{eq}{l}
e^{i\rvec p\rvec x}={\SUM_{L=0}^\infty} (1+2L)\ro P^L(\cos\th)i^Lj_L(Pr),~~
\ro P^L(\cos\th)=\sqrt{4\pi\over 1+2L}\ro Y^L_0(\phi,\th)
\end{eq}

Nonrelativistic  scattering theory is described with 
 the irreducible $\SU(2)\sx\R^3$-re\-pre\-sen\-ta\-tions.
The Hilbert spaces 
induced by a trivial or faithful re\-pre\-sen\-ta\-tion $\SO(2)$
on $W\cong\C^n$, $n=1,2$,
have a   measure related  distributive basis with generalized 
(distributive) eigenstates 
\begin{eq}{ll}
\hbox{for }J=0:&\{\rstate{P^2,0;\rvec \om,h}\mid\rvec\om\in\Om^2,~h=0\}\cr
\hbox{for }J={1\over2},1,\dots:&
\{\rstate{P^2,J;\rvec \om, h}\mid\rvec\om\in\Om^2, h =\pm 1\}
\end{eq}By abuse of language
since not a Hilbert space vector, an element of the distributive basis
$\rstate{P^2,J;\rvec\om, h}$ is called a scattering `eigenstate' with momentum 
$\rvec p$ of  absolute value $P$ (translation invariant),
  direction $\rvec\om$ (translation eigenvalues) 
and polarization $J=0,{1\over2},1,\dots$ 
(rotation invariant) with $\SO(2)$-eigenvalues $h$.
The notation for an `eigenstate' shows the re\-pre\-sen\-ta\-tion characterizing
invariants, here $(P^2,J)$, before the semicolon and the
eigenvalues in this re\-pre\-sen\-ta\-tions, here $(\rvec\om,h)$, behind the
semicolon.

The distributive basis is acted upon with the inducing 
$\SO(2)\x\R^3$-re\-pre\-sen\-ta\-tion
\begin{eq}{rl}
 (e^{\pm i\chi},\rvec x)\m\rstate{ P^2,J;\rvec \om, h}
&=  e^{ h iJ\chi}e^{-i P\rvec \om\rvec x}
\rstate{ P^2,J;\rvec \om, h}\cr
\end{eq}The momentum direction on the sphere ${\rvec p\over|\rvec p|}=\rvec \om\in\Om^2\cong\SU(2)/\SO(2)$
is the axis for the fixgroup $\SO(2)$ rotations
\begin{eq}{l}
\Om^2\ni{\rvec p\over|\rvec p|}=\rvec \om={\scriptsize\pmatrix{
\sin\th\cos\phi\cr
\sin\th\sin\phi\cr
\cos\th\cr}},~~ e^{-i\rvec p\rvec x}=e^{-i|\rvec p|\rvec\om\rvec x},~~
\int d^3 p=
 \int_0^\infty |\rvec p|^2 d|\rvec p|\int d^2\om\cr
 \int d^2\om=\int _0^{2\pi}d\phi
\int_{-1}^1d\cos\th,~~\de(\rvec \om)={1\over\sin\th}\de(\th)\de(\phi)
\cr
\end{eq}

The scalar product distribution is the 
product of the two scalar products for the
semidirect factors
\begin{eq}{rl}
\sprod{P^2,J; \rvec \om', h'}{P^2,J; \rvec \om, h}&=\de_{ h h'}
 \de({\rvec \om-\rvec \om'\over4\pi})\cr
 
 \int{d^2\om\over 4\pi}
\rstate{P^2,J;\rvec \om, h}\lstate{P^2,J;\rvec \om, h}
&\cong \bl 1_n\id_{L^2(\Om^2)}=\id_{L^2(\Om^2,\C^n)},~~n=1,2
\end{eq}With respect to the cyclic, but reducible position translation
re\-pre\-sen\-ta\-tions 
$\R^3\ni\rvec x\mape j_0(Pr)=\int d^2\om e^{-iP\rvec\om\rvec x}$
the scalar product involves the  positive and orthogonal  Dirac 
distribution 
on the 2-sphere $\de(\rvec\om-\rvec\om')$.
With respect to the fixgroup re\-pre\-sen\-ta\-tions 
$\SO(2)\map \SU(n)$
the Kronecker $\de_{ h h'}$ shows the scalar product in $\C^n$, $n=1,2$.

The Hilbert space consists of  2-sphere square integrable
momentum wave packets $L^2(\Om^2,\C^n)$ valued in $\C^n$, $n=1,2$
\begin{eq}{l}
f\in L^2(\Om^2,\C^n):~~
\left\{\begin{array}{rl}
\rstate{P^2,J;f}&
=\int {d^2\om\over4\pi}~f(\rvec \om)_h\rstate{P^2,J;\rvec\om, h}\cr
\sprod {P^2,J;f_2}{P^2,J;f_1}&
=\int {d^2\om\over4\pi}~\ol{f_2(\rvec \om)_h}f_1(\rvec \om)_h\cr
\end{array}\right.
\end{eq}The transformation behavior of the Hilbert space vectors 
is built by  that of the distributive basis.
The  $\SO(3)$-re\-pre\-sen\-ta\-tions are induced 
by the fixgroup $\SO(2)$-re\-pre\-sen\-ta\-tions\cite{S041}. 

All this can be seen as a distributional generalization
of the finite dimensional case, e.g. from one  basic vector
with time dependence $\rstate E\mape e^{iEt}\rstate E$
for the vector space $\C$
to a distributive basis $\{\rstate{P^2,J;\rvec\om, h}\}$ 
for the infinite dimensional  re\-pre\-sen\-ta\-tion space 
$L^2(\Om^2,\C^n)$.
There are cyclic normalized vectors
for the $\SU(2)\sx\R^3$-re\-pre\-sen\-ta\-tions, 
obtained by 2-sphere integration   
 with the constant function $1\in L^2(\Om^2)$, $f(\rvec\om)=1$
 all elements of a  distributive basis 
\begin{eq}{rl}
\rstate {P^2,J;1,h}&=
\int {d^2\om \over4\pi}
 \rstate{P^2,J;\rvec\om, h}\in L^2(\Om^2,\C^n)\cr
 \sprod{P^2,J;1,h'}{P^2,J;1,h}&=\de_{hh'}j_0(0)=\de_{hh'}\cr
\end{eq}The matrix element for time translation re\-pre\-sen\-ta\-tions
$\sprod{E,t_2}{E,t_1}=e^{iEt}$, $t=t_1-t_2$, has the analogue matrix element
for the Euclidean position group re\-pre\-sen\-ta\-tions
on this cyclic vector
\begin{eq}{rl}
\int {d^2 \om_1 d^2 \om_2\over(4\pi)^2} 
\sprod{P^2,J;\rvec \om_2, h_2 ,\rvec x_2}
{P^2,J;\rvec \om_1 , h_1,\rvec x_1}
&=
\de_{h_1h_2}\int {d^2\om\over4\pi} e^{-iP\rvec\om\rvec x}
\cr
= \sprod{P^2,J;1,h_2,\rvec x_2}{P^2,J;1,h_1  ,\rvec x_1}
&=\de_{h_1h_2}j_0(Pr),~~\rvec x=\rvec x_1-\rvec x_2
\cr
\end{eq}

Hilbert spaces for different  
translation or rotation invariant $\{P^2,L\}$ are orthogonal
as seen in the Schur scalar product for the re\-pre\-sen\-ta\-tion
matrix elements,
integrating with Haar measure $d^3 x$ over all translations 
\begin{eq}{l}
\{P^2,L\mid P^2>0,~L=0,1,\dots\}\hbox{ with }
D_m^{P^2,L}(\rvec x)=\sqrt{4\pi\over 1+2L}
\ro Y^L_m(\phi,\th) i^Lj_L(Pr) \cr
\begin{array}{rl}
\rsprod{D_{m_1}^{P^2_1,L_1}}{D_{m_2}^{P^2_2,L_2}}&=
\int d^3 x~\ol{D_{m_1}^{P_1^2,L_1}(\rvec x)}
D_{m_2}^{P_2^2,L_2}(\rvec x)\cr
&={1\over 1+2L_1}\de_{L_1L_2}\de_{m_1m_2}
\int_0^\infty  r^2 dr~ j_{L_1}(P_1r)j_{L_1}(P_2 r)\cr
&={1\over 1+2L_1}\de_{L_1L_2}
\de_{m_1m_2}{1\over 4P_1^2}\de({P_1-P_2\over 2\pi})\end{array}
\end{eq}

\subsection
[- for Stable Relativistic  Particles with Spacetime Translations]
{- for Stable Relativistic Particles\\ with Spacetime Translations}

The  re\-pre\-sen\-ta\-tions of  
the semidirect Poincar\'e group
with the orthochronous Lorentz group acting on the spacetime translations 
\begin{eq}{l}
\SL(\C^2)\sx\R^4,~~(\La,x)\in\SO_0(1,3)\sx\R^4\cr
(\La_1, x_1)\o (\La_2,x_2)=(\La_1\La_2,x_1+\La_1 x_2)\cr
\SO_0(1,3)\sx\R^4/\SO_0(1,3)
\cong\R^4\hbox{ (no group isomorphism)}
\end{eq}act - if nontrivial -  on an infinite dimensional Hilbert space. 
There is Wigner's classification\cite{WIG} of these re\-pre\-sen\-ta\-tions
according to the fixgroup types for energy-momenta.
 
The re\-pre\-sen\-ta\-tions for  a  stable massive particle
arise by straightforward Lorentz compatible composition of time and
position space structures:
The fixgroup of  energy-momenta $p\in\R^4$ with  $p^2=m^2>0$ is the
rest system related position  rotation group $\SO(3)$.
The Poincar\'e group re\-pre\-sen\-ta\-tion matrix elements collect the  
irreducible spacetime translation
re\-pre\-sen\-ta\-tions with eigenvalues on the forward and backward 
energy-momentum hyperboloids $\cl Y_\pm^3\cong\SO_0(1,3)/\SO(3)$, e.g. 
for the Lorentz scalar re\-pre\-sen\-ta\-tion  matrix elements, relevant, e.g., 
for a stable pion
\begin{eq}{rl}
 \R^4\ni  x\mape& 
\int{d^4 p\over 2\pi m^2}\de( p^2-m^2)e^{ipx}
={\vth(-x^2)2\cl K_1(|mx|)-\vth(x^2)\pi\cl N_{-1}(|mx|)\over
|mx|}\cr
\hbox{for }&m>0\hbox{ with }|x|=\sqrt{|x^2|}
\hbox{, supported by } p^2=m^2
\end{eq}Starting from flat spacetime $\R^2$ with
rotation free  Poincar\'e group $\SO_0(1,1)\sx\R^2$ and  
1-dimensional energy-momentum  hyperbolas 
$\cl Y^1_\pm\cong\SO_0(1,1)$  the 
embedding into 4-dimensional spacetime with 
 rotation $\SO(3)$ degrees of freedom 
can be obtained with  a 2-sphere spread by a 
Lorentz invariant derivative ${\p\over\p {x^2\over 4\pi}}$
\begin{eq}{rl}
\int d^4 p~\de( p^2-1)e^{ipx}
&={\p\over\p {x^2\over 4\pi}}\int d^2 p~ \de( p^2-1)e^{ipx}|_{x=(t,r)}\cr
&={\p\over\p {x^2\over 4\pi}}[\vth(-x^2)2\cl K_0(|x|)-\vth(x^2)\pi\cl N_{0}(|x|)]
\end{eq}The  Neumann functions\cite{GEL1,VIL}  $\cl N_n$ for timelike translations 
$\vth(x^2)$
and the Macdonald functions $\cl K_n$ for spacelike translations $\vth(-x^2)$
integrate  re\-pre\-sen\-ta\-tion matrix elements of 1-dimensional 
translations over the hyberboloid
\begin{eq}{rrl}
\R\ni t\mape& -\pi \cl N_0(t)&=\int d\psi~\cos (t\cosh\psi)\cr
\R\ni r\mape& 
2\cl K_0(r)&=\int d\psi~e^{-|r|\cosh\psi}\cr
 \end{eq}Matrix elements of nontrivial re\-pre\-sen\-ta\-tions of the Lorentz
group $\SO_0(1,3)$ are obtained by translation derivations 
${\p\over\p x}=2x{\p\over\p x^2}$.

The Hilbert space 
for a stable massive particle with $\SL(\C^2)\sx\R^4$-re\-pre\-sen\-ta\-tion,
induced by a  finite dimensional
 re\-pre\-sen\-ta\-tion of  the  spin-translation group
on $W\cong\C^{1+2J}$
\begin{eq}{rl}
\SU(2)\x\R^4&\map \U(1+2J)\cr
(u,x)&\mape 
 2J(u)e^{ipx}\hbox{ with }p^2=m^2
\end{eq}has a distributive basis on the forward energy-momentum hyperboloid
\begin{eq}{l}
\{\rstate{m^2,J;\rvec p,a}=\rstate{m^2,J;\ty c,a}\mid \rvec
p\in\R^3,~\ty c\in\cl Y^3,a=1,\dots,1+2J\}\cr 
\end{eq}The hyperboloid  is parametrizable with  hyperbolic coordinates 
$\ty c$,  appropriate for the Lorentz group action, 
or with the more familiar momentum
coordinates $\rvec p$ 
\begin{eq}{l}
\cl Y_\pm^3\ni \vth(p^2)\vth(\pm p_0){p\over|p|}=\pm\ty c =
{\scriptsize\pmatrix{
\pm\cosh\psi\cr
{\rvec p\over |\rvec p |} \sinh\psi\cr
}}
={1\over |p|}{\scriptsize\pmatrix{
\pm p_0 \cr
\rvec p\cr}},~~p_0=\sqrt{|p|^2+\rvec p^2}
\cr
\vth(p^2)\vth(\pm p_0)e^{ipx}=e^{\pm i|p|\ty c x},~~\int d^4 p~\vth(\pm p_0)\vth(p^2)=
 \int_0^\infty |p|^3 d|p|\int_\pm d^3\ty c\cr
  \int_+ d^3\ty c=\int_{0}^\infty (\sinh\psi)^2 d\psi \int d^2\om
 = \int {d^3p\over 2p_0},~~
\de(\ty c)
={1\over(\sinh\psi)^2}\de(\psi)\de(\rvec \om)\cr

\end{eq}The elements of the distributive basis
$\rstate{m^2,J;\rvec p,a}$ (no Hilbert space vectors) are called  
`eigenstates'  for a particle with the invariants
mass $m$, spin $J$  and  eigenvalues momentum $\rvec p$ and 3rd spin component $a$.


The scalar product distribution for the `eigenstates'
is the product of the scalar product in the finite dimensional spin
re\-pre\-sen\-ta\-tion space and
the  Schur scalar product  for the translations
with the Dirac distribution on the energy-momentum hyberboloid
\begin{eq}{rl}
\sprod {m^2,J;\rvec p_2,a_2}{m^2,J;\rvec p_1,a_1}&=
\de_{a_1a_2} 8\pi p_0~\de(\rvec p_1-\rvec p_2)\cr
=\sprod {m^2,J;\ty c_2,a_2}{m^2,J;\ty c_1,a_1}&=
\de_{a_1a_2}
4\pi\de(\ty c_1-\ty c_2)\cr
\end{eq}The re\-pre\-sen\-ta\-tion space is the Hilbert space 
of the square integrable mappings on the forward 
energy-momentum hyperboloid. 
The completeness reads
\begin{eq}{rl}
\int{d^3 p\over 8\pi p_0}
\rstate {m^2,J;\rvec p,a}\lstate{m^2,J;\rvec p,a}
&=\int_+{d^3\ty c\over 4\pi}
\rstate {m^2,J;\ep,\ty c,a}\lstate{m^2,J;\ep,\ty c,a}
\cr
&\cong\bl 1_{1+2J}\id_{L^2(\cl Y^3)}= \id_{L^2(\cl Y^3,\C^{1+2J})}
\end{eq}

The Hilbert space vectors
$\rstate{m^2,J;f}$ 
use an expansion with square integrable functions of
 the hyperbolic energy-momentum directions 
\begin{eq}{l}
f\in  L^2(\cl Y^3,\C^{1+2J}):\left\{\begin{array}{rl}
\rstate{m^2,J;f}
&=\int {d^3 p\over 8\pi p_0}~f(\rvec p)_a
\rstate{m^2,J;\rvec p,a}\cr
&=\int_+ {d^3\ty c\over 4\pi}~f(\ty c)_a
\rstate{m^2,J;\ty c,a}\cr
\sprod {m^2,J;f_2}{m^2,J;f_1}
&=\int {d^3 p\over 8\pi p_0}
~\ol{f_2(\rvec p)_a}f_1(\rvec p)_a\cr
&=\int_+ {d^3\ty c\over4\pi}~\ol{f_2(\ty c)_a}f_1(\ty c)_a\cr
\end{array}\right.
\end{eq}The spacetime translation action on the Hilbert space vectors 
is built by  that on the distributive basis.
The Lorentz $\SL(\C^2)$-re\-pre\-sen\-ta\-tions  are induced by the rest system
spin $\SU(2)$-re\-pre\-sen\-ta\-tions\cite{WIG,S041}.

In contrast to  position space with compact homogeneous rotation group
$\SU(2)$, the re\-pre\-sen\-ta\-tion matrix elements of the Poincar\'e group 
with noncompact homogeneous Lorentz group $\SL(\C^2)$ are not square
integrable. The normalization of the integral of the basis
distribution with the constant function $f(\ty c)=1$ is  no element 
of $L^2(\cl Y^3,\C^{1+2J})$ because of the infinite hyperboloid volume
\begin{eq}{l}
\rstate{m^2,J;1,a}=\int_+ {d^3\ty c\over 4\pi} \rstate{m^2,J;\ty c,a }
\notin L^2(\cl Y^3,\C^{1+2J})\cr
\sprod{m^2,J;1,a_1}{m^2,J;1,a_2}=\de_{a_1a_2}
\int {d^4p\over 2\pi}\vth(p_0)\de(p^2-1)\cr
\end{eq}Its translation dependent scalar product involves an 
 $\SO(1,3)\sx\R^4$-re\-pre\-sen\-ta\-tion matrix element
\begin{eq}{rl} 
\rstate{m^2,J;1,a,x}
=\int_+ {d^3\ty c\over 4\pi}
 \rstate{m^2,J;\ty c,a , x}\cr
\int_+ {d^3\ty c_1 d^2\ty c_2\over(4\pi)^2} \sprod{m^2,J;\ty c_2 ,a_2, x_2}
{m^2,J;\ty c_1,a_1 , x_1}
&=\de_{a_1a_2}\int_+ {d^3\ty c\over4\pi} e^{im \ty c x},~~x= x_1-x_2\cr
=\sprod{m^2,J;1,a_2,x_2}{m^2,J;1,a_1, x_1}
&=\de_{a_1a_2}\int{d^4 p\over2\pi}\vth(p_0)\de(p^2-1)e^{i p x}\cr
\end{eq}

The projections\cite{BS03} to the nonrelativistic time and position 
`states' is effected 
by the appropriate Dirac distributions
\begin{eq}{rl}
\int{d^4p\over2\pi m^2}\de(\rvec p-\rvec k)\de(p^2-m^2) e^{ipx}
&=\int{dp_0\over2\pi m^2}
\de(p_0^2-m^2-\rvec k^2) e^{ip_0t-i\rvec k\rvec x}\cr
&={1 \over2\pi m^2}{\cos k_0\over k_0} t~e^{-i\rvec k\rvec x}  
,~~k_0=\sqrt{m^2+\rvec k^2}
\cr
\hbox{for rest system }\rvec k=0:&
={\cos m t \over2\pi m^3}\cr 
\cr

\int{d^4p\over2\pi m^2}\de(p_0-E)\de(p^2-m^2) e^{ipx}
&=\int{d^3p\over2\pi m^2}~
\de(E^2-\rvec p^2-m^2) e^{iEt-i\rvec p\rvec x}\cr

&=e^{iEt}\vth(P^2){P\over m^2}{\sin Pr\over P r},~~P=\sqrt {E^2-m^2}

\end{eq}

Hilbert spaces for stable particles $\{m^2,J\}$ with different 
translation or rotation invariant (or also 
with different internal charge 
$z\in\Z$, e.g. for   particle-antiparticle)  
are orthogonal. In the corresponding Schur scalar product 
there arises the infinite volume of the
energy-momentum hyperboloid 
\begin{eq}{rl}
D^{m^2}(x)&=\int{d^4p\over 2\pi m^2}\de(p^2-m^2) e^{ipx}\cr
\rsprod{D^{m^2_1}}{D^{m^2_2}}&=
\int {d^4 x\over 4\pi}~\ol{D^{m_1^2}(x)}D^{m_2^2}(x)
={1\over 4m^3_1}\de({m_1-m_2\over 2\pi})\int d^4p~\de(p^2-1)
\end{eq}

\subsection{Selfdual Translation Re\-pre\-sen\-ta\-tions}

Both for position and spacetime translation re\-pre\-sen\-ta\-tions,
there arise the trigonometric functions, as seen,  e.g., 
in $t\mape (\cos \mu t,\sin \mu t)$ as matrix
elements of selfdual time re\-pre\-sen\-ta\-tions.
They combine dual  re\-pre\-sen\-ta\-tions\cite{LIE13} of time translations 
\begin{eq}{l}t\mape e^{\pm i\mu t} 
\end{eq}which act upon a  2-dimensional re\-pre\-sen\-ta\-tion space 
$\C\rstate{\mu }\pl\C\lstate{\mu }\cong\C^2$ as the direct sum 
spanned by eigenvector (`ket')
and dual eigenvector (`bra').
Such a selfdual time translation re\-pre\-sen\-ta\-tion is used, e.g.,
 in the position-momentum formulation 
 of the harmonic oscillator with time development
\begin{eq}{rl}
{\scriptsize\pmatrix{e^{i\mu t}&0\cr 0&e^{-i\mu t}\cr}}
&= \int d E~{\scriptsize\pmatrix{\de(E-\mu)&0\cr 0&\de(E+\mu)\cr}} e^{iEt}\cr

\cong 
{\scriptsize\pmatrix{\cos \mu t&i\sin \mu t\cr i\sin \mu t&\cos \mu t\cr}} 
&= \int d E~\ep(\mu){\scriptsize\pmatrix{\mu&E\cr E&\mu\cr}}\de(E^2-\mu^2) 
e^{i Et}\cr
\hbox{ supported by}& E=\pm \mu

\end{eq}where position and momentum are real 
and imaginary part $(\bl x,-i\bl p)={\ro u\pm\ro u^\star\over\sqrt
2}$ of creation and annihilation operators with dual time development.

The basis  distribution for
forward and backward energy-momentum hyperboloid
with  creation and annihilation operators
for the Poincar\'e group,
acted upon with  dual translation re\-pre\-sen\-ta\-tions, 
can be seen explicitly in a field expansion, e.g. for a stable neutral pion 
\begin{eq}{l}
\bl\Phi(x)
=\int {d^3p\over 4\pi p_0 m}
{  e^{ipx}\ro u(\rvec p)+ e^{-ipx}\ro u^\star(\rvec p)\over \sqrt 2}
\hbox{ with }p_0=\sqrt{m^2+\rvec p^2}\cr
\end{eq}The distributive basis has the 
scalar product distribution, written as Fock state expectation value 
$\angle{...}$
\begin{eq}{l}
\angle{\ro u^\star(\rvec p_2)\ro u(\rvec p_1)}
=8\pi p_0\de(\rvec p_1-\rvec p_2),~~
\angle{\ro u(\rvec p_2)\ro u(\rvec p_1)}
=\angle{\ro u^\star(\rvec p_1)\ro u^\star(\rvec p_2)}=0
\end{eq}which leads to the scalar Poincar\'e group  re\-pre\-sen\-ta\-tion matrix element
as Fock value  of the anticommutator
\begin{eq}{l}
\angle{\acom{\bl\Phi(x_2)}{\bl\Phi(x_1)}}
=\int{d^4p\over2\pi m^2}\de(p^2-m^2) e^{ipx},~~x=x_1-x_2\cr
\end{eq}

The Feynman propagator contains, 
in addition to the Fock value of the anticommutator,
the quantization commutator multiplied with the causal order function
$\ep(x_0)$ 
\begin{eq}{l}\angle{\acom{\bl\Phi(x_2)}{\bl\Phi(x_1)}-\ep(x_0)
\com{\bl\Phi(x_2)}{\bl\Phi(x_1)}}
=-\int{d^4p\over2i\pi^2 m^2}{1\over p^2+io -m^2} e^{ipx}\cr
\ep(x_0){\com{\bl\Phi(x_2)}{\bl\Phi(x_1)}}
=-\int{d^4p\over2i\pi^2 m^2}{1\over p^2_\ro P -m^2} e^{ipx}
\cr\end{eq}The $\ep(x_0)$-multiplied quantization is no 
re\-pre\-sen\-ta\-tion matrix element of the Poincar\'e group\cite{S96}
as visible in
the off-shell contributions ($\ro P$ denotes the principal value integration).


\section{The Hilbert Spaces for Unstable Particles}

Unstable particles are described with complex translation invariants,
starting with $\mu+i\Ga$ for an energy $\mu$ with a nontrivial width  $\Ga>0$
 for time translations. 
All re\-pre\-sen\-ta\-tions  for  unstable states (particles), by abuse of language called
unstable re\-pre\-sen\-ta\-tions, are infinite
dimensional, even for abelian time translations.
For a first survey, only spinless unstable states (particles) are considered.

\subsection
{- for Unstable Energy States}

An irreducible   re\-pre\-sen\-ta\-tion matrix element
of time translations can be written with a Dirac energy distribution
supported by the translation invariant $\mu\in\R$, i.e. as a residual
re\-pre\-sen\-ta\-tion\cite{S041}
\begin{eq}{l}
\R\ni t\mape e^{i\mu t}
=\int dE~\de(E-\mu)e^{iEt}
=\oint{dE\over2i\pi}{1\over E-\mu}e^{iEt}\cr
\hbox{supported by }E=\mu\cr
\end{eq}

Finite dimensional time translation re\-pre\-sen\-ta\-tions with nontrivial 
width have to be indefinite unitary . They start with
the 2-dimensional re\-pre\-sen\-ta\-tions 
\cite{S012} involving complex conjugate energies
\begin{eq}{l}
\R\ni t\mape e^{i\mu t}{\scriptsize\pmatrix{e^{-\Ga t}&0\cr0&e^{\Ga t}\cr}}
\in\U(1)\x\SO_0(1,1)\subnoteq \U(1,1)
\end{eq}The sum of the two matrix elements 
multiplied with the characteristic functions for future and past
is a matrix element of an unstable Hilbert re\-pre\-sen\-ta\-tion 
\begin{eq}{rl}
\R\ni t&\mape  \vth(t)e^{i(\mu+i\Ga)t}
+\vth(-t)e^{i(\mu-i\Ga)t}=
e^{i\mu t-\Ga |t|}\cr
&=\int{dE\over\pi}
{\Ga\over
(E-\mu)^2+\Ga^2}e^{iEt}=D^{\mu,\Ga}(t) \hbox{ with }\mu\in \R,~\Ga\ge 0
\end{eq}normalized for the neutral time translation $t=0$
(present).
The spectral function for the energies in the
 matrix element of unstable time translation 
re\-pre\-sen\-ta\-tions
\begin{eq}{l}
\int{dE\over\pi}~{\Ga\over (E-\mu )^2+\Ga^2}e^{iEt}
=\oint{dE\over 2i\pi}[{\vth(t)\over E-\mu -i\Ga}-{\vth(-t)\over E-\mu +i\Ga}]
e^{iEt}\cr
\hbox{supported by }E=\mu\pm i\Ga

\end{eq}has two complex conjugated poles
as invariants. It is a Breit-Wigner function (`$\Ga$-widened  Dirac distribution').
The Dirac distribution is 
the imaginary part of the  advanced and retarded distribution 
\begin{eq}{l}
\de(E-\mu )={1\over 2i\pi}[{1\over E-\mu -io}-{1\over E-\mu +io}]=
{1\over\pi}{o\over (E-\mu )^2+o^2}
\hbox{ with positive }o\to 0
\end{eq}

An unstable time translation re\-pre\-sen\-ta\-tion
for $\Ga>0$  is  a cyclic, but reducible time translation re\-pre\-sen\-ta\-tion.
It  is the direct integral  over 
irreducible re\-pre\-sen\-ta\-tions $t\mape e^{iEt}$ for all energies $E\in\R$,
distributed with the positive function 
$E\mape{1\over\pi}~{\Ga\over (E-\mu )^2+\Ga^2}$.
An unstable time re\-pre\-sen\-ta\-tion 
is an irreducible re\-pre\-sen\-ta\-tion of the semidirect
product of the discrete time reflection group acting upon the time translations
\begin{eq}{l}
(\ep,t)\in \O(1)\sx\R,~~\ep\in \{\pm1\}=\O(1)\cr
(\ep_1,t_1)\o(\ep_2,t_2)= (\ep_1\ep_2, t_1+\ep_1t_2)\cr
\O(1)\sx\R/\O(1)\cong\R
\hbox{ (no group isomorphism)}
\end{eq}Faithful re\-pre\-sen\-ta\-tions of $\O(1)\sx\R$
are inducable  by  $\R$-re\-pre\-sen\-ta\-tions with trivial fixgroup. 
Irreducible reflection group re\-pre\-sen\-ta\-tions are 1-dimensional - either 
invariant (trivial) under time reflection 
$\rstate +\stackrel{-1}\mape \rstate +$ or faithful $\rstate -\stackrel{-1}
\mape -\rstate -$.

The Hilbert space for an unstable state
is infinite dimensional. It has a distributive basis involving all energies
\begin{eq}{l} 
\Ga>0:~~\{\rstate{\mu ,\Ga;E}\mid E\in\R\}
,~~\rstate{\mu ,\Ga;E,t}=e^{iEt}\rstate{\mu ,\Ga;E}
\end{eq}By abuse of language $\rstate {\mu ,\Ga;E}$  is called an unstable `eigenstate'
with time translation eigenvalue $E$ for the  invariant $\mu $
with width $\Ga$.
The Hilbert space vectors for the unstable time translation re\-pre\-sen\-ta\-tions
\begin{eq}{l}
\rstate{\mu ,\Ga;f}=\int
dE~f(E)\rstate{\mu ,\Ga;E},~~\rstate{\mu ,\Ga;f,t}=\int
dE~f(E)\rstate{\mu ,\Ga;E,t} \cr
\end{eq}are defined with Fourier transformable elements 
$ f(E)=\int dt~ \tilde f(t)e^{iEt}$ 
from  the convolution Banach algebra $L^1(\R)$
(group algebra) with the 
absolute integrable function classes $\int dt~|\tilde f(t)|<\infty$
on the time translation group.
Their scalar product for $\Ga>0$  is  not defined by square integrability,
but by a positive type function\cite{GELRAI,FOL} (more in the next section)
from the group algebra dual $L^\infty(\R)$, the essentially bounded 
function classes $\Sup_t|\tilde f(t)|< \infty$.
The unstable state scalar product is defined  exactly by the  matrix element 
$\{t\mape D^{\mu ,\Ga}(t)\}\in L^\infty(\R)$
of the cyclic time translation re\-pre\-sen\-ta\-tion 
\begin{eq}{rl}
\sprod{\mu ,\Ga;f_2}{\mu ,\Ga;f_1}
&=\int{dE\over \pi}~\ol{f_2(E)}~{\Ga\over (E-\mu )^2+\Ga^2}~f_1(E)\cr
&=\int dt_1dt_2~\ol{\tilde f_2(t_2)}~
D^{\mu ,\Ga}(t_1-t_2)~\tilde f_1(t_1)\cr
\end{eq}The Hilbert space vectors are the  finite norm functions 
$\sprod{\mu ,\Ga;f}{\mu ,\Ga;f}<\infty$.
The scalar product 
involves the positive scalar product distribution for the distributive basis
\begin{eq}{l}
\sprod{\mu ,\Ga;E'}{\mu ,\Ga;E}
={1\over\pi}~{\Ga\over (E-\mu )^2+\Ga^2}\de({E-E'\over 2\pi}) 

\end{eq}

It is important to realize that - according to Gel'fand and Raikov
\cite{GELRAI} -
cyclic re\-pre\-sen\-ta\-tions  and scalar product inducing positive
type functions are uniquely related to each other.

Integrating with the constant function $f(E)=1$ over all energies,  one obtains 
a cyclic vector for the unstable time re\-pre\-sen\-ta\-tion
\begin{eq}{rlrl}
\rstate{\mu ,\Ga;1}&=\int dE~\rstate{\mu ,\Ga;E},&\sprod{\mu ,\Ga;1}{\mu
,\Ga;1}&=1\cr
\rstate{\mu ,\Ga;1,t}&=\int dE~e^{iEt}\rstate{\mu ,\Ga;E},&
\sprod{\mu ,\Ga;1}{\mu ,\Ga;1,t}
&=D^{\mu ,\Ga}(t)\cr
\end{eq}The associated Fourier transform 
$\int{dE\over2\pi}1e^{-iEt}=\de(t)=\de_0(t)$ is no element of the time translation group
algebra $\de_0\notin L^1(\R)$. The distribution $\de_0$ embeds the 
trivial time translation  $t=0$. It can be approximated 
by $L^1(\R)$-elements (approximate identity\cite{FOL}).

The stable case for the Hilbert space functions 
is rediscovered as limit $\Ga\to 0$.
It reduces  the distributive eigenvector basis for all energies $E\in\R$ to 
one basic vector for one energy
$E=\mu $ 
\begin{eq}{rl}
\Ga\to0:~~\sprod{\mu ,\Ga;f_2}{\mu ,\Ga;f_1}
&\to\int dE~\ol{f_2(E)}~\de(E-\mu )~f_1(E)\cr
&=\int dt_1dt_2~\ol{\tilde f_2(t_2)}~
e^{i\mu (t_1-t_2)}~\tilde f_1(t_1)\cr
&=\ol{f_2(\mu )}~f_1(\mu )\hbox{ with }\rstate{\mu ;f}=f(\mu )\rstate \mu \cr

\end{eq}

The Schur product for the unstable time 
translation re\-pre\-sen\-ta\-tions
\begin{eq}{l}
\{D^{\mu ,\Ga}\mid \mu \in\R,~~\Ga\ge0\}
\hbox{ with }D^{\mu ,\Ga}(t)=e^{i\mu t-\Ga|t|}\cr
\rsprod{D^{\mu _2,\Ga_2}}{D^{\mu _1,\Ga_1}}=
\int dt~e^{i(\mu _1-\mu _2)t-(\Ga_1+\Ga_2)|t|}=
2{\Ga_1+\Ga_2\over (\mu _1-\mu_2)^2+(\Ga_1+\Ga_2)^2}
\end{eq}is not orthogonal. It gives nontrivial transition elements for unstable re\-pre\-sen\-ta\-tions. 
Unstable particles can constitute collectives with other particles - stable or
unstable (more below).

\subsection
{- for Unstable Scattering States}

The transition from stable to unstable scattering states 
in position space $\R^3$ with complex translation invariant is obtained
by widening - with width $\ga={\Ga\over2}$ - the
Dirac distribution on the momentum 2-sphere to a Breit-Wigner function
\begin{eq}{rl}
\de(\rvec p^2-P^2)\to 
  \de_\ga(\rvec p^2-P^2)&=
{1\over2i \pi}
[{1\over \rvec p^2-(P+i\ga)^2}-{1\over \rvec p^2-(P-i\ga)^2}]\cr
&={1\over\pi}{2P\ga\over (\rvec p^2-P^2+\ga^2)^2+4P^2\ga^2},~~P\ga>0 
\cr
\hbox{selfdual invariant singularities at } |\rvec p|&=\pm (P\pm i\ga)\cr
\end{eq}The unstable re\-pre\-sen\-ta\-tions of the Euclidean position  group
involve re\-pre\-sen\-ta\-tions of  
the position space reflection group  
\begin{eq}{l}
\O(3)\sx\R^3,~~
\O(3)\cong \O(1)\x \SO(3)\cr
\O(3)\sx\R^3/\O(3)\cong \R^3 
\hbox{ (no group isomorphism)}
\end{eq}

It has the scalar matrix elements, normalized for 
the neutral position translation $\rvec x=0$   
\begin{eq}{rl}
\R^3\ni \rvec x&\mape
\int{d^3 p\over2 \pi P}  \de_\ga(\rvec p^2-P^2)
e^{-i\rvec p\rvec x}\cr
&=-{1\over Pr}{\p\over\p r}
\int dp~ \de_\ga( p^2-P^2)
e^{-ipr}=-{1\over P r}{\p\over\p r}
[{e^{iPr}\over 2(P+i\ga)}+{e^{-iPr}\over 2(P-i\ga)}]e^{-\ga r}\cr
&={\sin Pr\over Pr}
e^{-\ga r}=D^{P^2,\ga^2}(\rvec x)\hbox{ with } P>0,~\ga\ge0
\cr
\end{eq}

The distributive basis for the expansion of the
Hilbert space vectors from $L^1(\R^3)$ uses all momenta
\begin{eq}{rlrl} 
\ga>0:&\{\rstate{P^2,\ga;\rvec p}\mid \rvec p\in\R^3\},&
\rstate{P^2,\ga;\rvec p,\rvec x}
&=e^{-i\rvec p\rvec x}\rstate{P^2,\ga;\rvec p}\cr
&\rstate{P^2,\ga;f}=\int {d^3p\over 2\pi}~f(\rvec p)\rstate{P^2,\ga;\rvec p},&
\rstate{P^2,\ga;f,\rvec x}&=
\int {d^3p\over 2\pi}~f(\rvec p)\rstate{P^2,\ga;\rvec p,\rvec x}
\end{eq}Again, $\rstate {P^2,\ga;\rvec p}$  
is called an unstable scattering `eigenstate'
with position translation eigenvalue $\rvec p$ for the invariant 
$P^2$ with width $\ga^2$.

The instability characteristic structure is 
the scalar product which  is not defined by square integrability, 
but exactly by the positive  momentum spectral function which occurs 
in the scalar re\-pre\-sen\-ta\-tion matrix elements
 \begin{eq}{rl}
\sprod{P^2,\ga;\rvec p_2}{P^2,\ga;\rvec p_1}
&={1\over P} \de_\ga(\rvec p_1^2-P^2)
2\pi\de(\rvec p_1-\rvec p_2)\cr
\sprod{P^2,\ga;f_2}{P^2,\ga;f_1}
&=\int{d^3p\over 2\pi P}~\ol{f_2(\rvec p)}~
 \de_\ga(\rvec p^2-P^2)
~f_1(\rvec p)\cr
\end{eq}The scalar product can also be written with the  positive type function
in the position translation parametrization
\begin{eq}{rl}
\sprod{P^2,\ga;f_2}{P^2,\ga;f_1}
&=\int d^3x_1d^3x_2~\ol{\tilde f_2(\rvec x_2)}~
D^{P^2,\ga^2}(\rvec x_1-\rvec x_2)
~\tilde f_1(\rvec x_1)\cr
\hbox{for } f(\rvec p)&=\int d^3x~ \tilde f(\rvec x)e^{-i\rvec p\rvec x}
\end{eq} 

A cyclic vector can be constructed  with an approximate identity
in $L^1(\R^3)$
\begin{eq}{l}
\rstate{P^2,\ga;1}=\int{d^3p\over 2\pi}~\rstate{P^2,\ga;\rvec p}
,~~\rstate{P^2,\ga;1,\rvec x}=
\int{d^3p\over 2\pi}e^{-i\rvec p\rvec x}\rstate{P^2,\ga;\rvec p}
\cr
\sprod{P^2,\ga;1}{P^2,\ga;1,\rvec x}=D^{P^2,\ga^2}(\rvec x)
\end{eq} 

The scalar product defining Breit-Wigner function 
$\rvec p \mape  \de_\ga(\rvec p^2-P^2)$ describes a $\ga$-shell around the
momentum 2-sphere with radius $P$.
For the stable case $\ga=0$ the support of the   Hilbert space functions 
is restricted
to the momentum 2-sphere. In this limiting case, i.e. for stable states,
the  functions 
have to be square integrable  
\begin{eq}{rl}
\ga\to 0:~~\sprod{P^2,\ga;f_2}{P^2,\ga ;f_1}
&\to \int{d^3p\over 2\pi P} ~\ol{f_2(\rvec p)}~
\de(\rvec p^2-P^2)
~f_1(\rvec p)\cr
&=\int{d^2\om\over 4\pi}~\ol{f_2(P,\rvec \om)}f_1(P,\rvec \om)\cr
&=\int d^3x_1d^3x_2~\ol{\tilde f_2(\rvec x_2)}~
{\sin P|\rvec x_1-\rvec x_2|\over P|\rvec x_1-\rvec x_2|}
~\tilde f_1(\rvec x_1)\cr
\end{eq}

The change from square integrable translation $\R^n$-re\-pre\-sen\-ta\-tions 
using the full (energy-)momentum space $\R^n$ 
and Plancherel unitarity
\begin{eq}{l}
f\in L^2(\R^n):~~\int{d^np\over (2\pi)^n}~\ol{f_2(p)}
f_1(p)
=\int d^nx~\ol{\tilde f_2( x_2)}\tilde f_1(x_1)
\end{eq}to stable re\-pre\-sen\-ta\-tions  
of the Euclidean group $\SU(2)\sx\R^3$ shows up in the 
restriction to the momentum 2-sphere $L^2(\Om^2)$ and the position spherical  
Bessel function. For unstable re\-pre\-sen\-ta\-tions
there is the  restriction with $e^{-\ga r}$  to $L^1(\R^3)$-
functions. 

\subsection
{- for Unstable Relativistic Particles}

The embedding of unstable time and position `states' into a Lorentz compatible
spacetime framework uses 
the Poincare group with reflection 
\begin{eq}{l}
\SO(1,3)\sx  \R^4,~~\SO(1,3)\cong\O(1)\sx\SO_0(1,3)\cr
\SO(1,3)\sx  \R^4/\SO(1,3)\cong\R^4
\hbox{ (no group isomorphism)}
\end{eq}It contains the groups $\O(1)\sx\R$ for time and $\O(3)\sx\R^3$ for space
translations.
The factor acting upon the spacetime translations
is the special Lorentz group which is itself a semidirect group
with  a 2-elementic reflection 
 group $\O(1)$ acting on the orthochronous Lorentz group
\begin{eq}{l}
(\ep,\La)\in\SO(1,3):~~(\ep_1,\La_1)\o(\ep_2,\La_2)
=(\ep_1\ep_2,\La_1\ep_1\La_2\ep_1)
\end{eq}The reflection group is not Lorentz invariant. It can be
realized e.g. with time reflections $\ep\cong{\scriptsize\pmatrix{-1&0\cr0&\bl1_3\cr}}$ 
or with space reflections $\ep\cong{\scriptsize\pmatrix{1&0\cr0&-\bl1_3\cr}}$.

The transition from stable to unstable particles
is obtained by  widening  the
Dirac distribution
on the backward and forward energy-momentum hyperboloid
to a Breit-Wigner function with a width $\ga={\Ga\over2}$
\begin{eq}{rl}
\de( p^2-m^2)\to  \de_\ga(p^2-m^2)&=
{1\over2i \pi}
[{1\over  p^2-(m+i\ga)^2}-{1\over p^2-(m-i\ga)^2}]\cr
&={1\over\pi}{2m\ga\over (p^2-m^2+\ga^2)^2+4m^2\ga^2},~~m\ga>0 
\cr
\end{eq}Therewith the unstable scalar Poincar\'e group
re\-pre\-sen\-ta\-tion matrix elements  are
\begin{eq}{rll}
\R^4\ni  x\mape&
\int{d^4 p\over2 \pi m^2}  \de_\ga(p^2-m^2)
e^{i p x}\cr
&=2{\p\over\p m^2 x^2}
\int d^2 p~   \de_\ga(p^2-m^2)
e^{i p x}|_{x=(t,r)}\cr
&=D^{m^2,\ga^2}(x) \hbox{ with } m>0,~\ga\ge0

\end{eq}

Since no longer only  `on-shell' $\de(p^2-m^2)$, 
the Hilbert space vectors need an expansion  with a distributive basis for
all energy-momenta
\begin{eq}{rlrl}
\ga>0:&\{\rstate{m^2,\ga;p}\mid p\in \R^4\},
&\rstate{m^2,\ga; p,x}&=e^{i p x}\rstate{m^2,\ga; p}\cr
&\rstate{m^2,\ga;f}=\int 
{d^4 p\over 2\pi}~ f(p)
\rstate{m^2,\ga;p},&
\rstate{m^2,\ga;f, x}&=
\int {d^4p\over 2\pi}~f( p)\rstate{m^2,\ga; p, x}
\end{eq}Again, $\rstate {m^2,\ga;p}$  
is called an unstable particle  `eigenstate'
with spacetime translation eigenvalue $p$ for the  invariant 
$m^2$ with width $\ga^2$.

The characterizing scalar product of 
the energy-momentum functions for an  unstable particle 
is defined exactly by the positive
energy-momentum  spectral function of the re\-pre\-sen\-ta\-tion. It
formalizes a $\ga$-shell around the stable
particle mass forward and backward hyperboloid  $\cl Y_\pm^3(m)$. Only for the stable limiting case
the Hilbert space vectors are square integrable functions supported by
the two shell energy-momentum hyperboloid 
\begin{eq}{rl}
\sprod{m^2,\ga;p_2}{m^2,\ga;p_1}
&={1\over m^2} \de_\ga(p_1^2-m^2)2\pi\de(p_1-p_2)\cr
\sprod{m^2,\ga;f_1}{m^2,\ga;f_2}
&=\int {d^4p\over 2\pi m^2}~\ol{f_2(p)}~ \de_\ga(p^2-m^2)f_1(p)\cr
\hbox{for }\ga\to 0:~~~&\to
\int {d^4p\over 2\pi m^2}~\ol{f_2(p)}~\de(p^2-m^2)f_1(p)\cr
&=\int {d^3p\over 4\pi p_0}~\ol{f_2(\rvec p)}f_1(\rvec p)\cr
\end{eq}

The projection\cite{BS03} on matrix elements for unstable energy `states' 
defines sharp momenta systems  
\begin{eq}{rl}
\int{d^4 p\over 2 \pi m^2}\de(\rvec p-\rvec k)  \de_\ga(p^2-m^2)
e^{i p x}
&=
{1\over 2\pi m^2 }\int d p~ \de_{\ga_k}(p^2-E^2)
e^{i p t-i\rvec k\rvec x}\cr
&={1\over 2\pi m^2 }
[{e^{iEt}\over 2(E+i\ga_k)}+{e^{-iEt}\over 2(E-i\ga_k)}] e^  {-\ga_k|t|}
 e^{-i\rvec k\rvec x}\cr
&={1\over 2\pi m^2 }
{E\cos Et+\ga_k\sin E t\over E^2+\ga_k^2} e^  {-\ga_k|t|}e^{-i\rvec k\rvec x} \cr
\rvec k=0\then E+i\ga_k=m+i\ga:~~~&= {1\over 2\pi m^2}
{m\cos mt+\ga\sin E t\over m^2+\ga^2} e^  {-\ga|t|} \cr

\end{eq}where the complex invariants $(E\pm i\ga_k)^2$
 arise from
 \begin{eq}{rl}
p_0^2-\rvec k^2-(m+i\ga)^2&=p_0^2-(E+i\ga_k)^2\cr
\then E+i\ga_k&=\sqrt{
\rvec k^2+m^2-\ga^2+2im\ga}\cr
&=\sqrt{ \rvec k^2+m^2}(1+{im\ga\over \rvec k^2+m^2}+\dots)\cr
\end{eq}The projection on matrix elements for unstable scattering `states' 
defines  sharp energy systems
\begin{eq}{rl}
\int{d^4 p\over2 \pi m^2}\de( p_0-E)  \de_\ga(p^2-m^2)
e^{i p x}
&=-
{P\over m^2}\int{d^3 p\over2 \pi P} \de_{\ga_E}(\rvec p^2-P^2)
e^{iEt-i\rvec p \rvec x}\cr
&=-e^{iEt}\vth(P^2){P\over m^2}{\sin Pr\over Pr}e^{-\ga_Er}\cr
\end{eq}where the complex invariants $(P\pm i\ga_E)^2$ arise from
\begin{eq}{rl}
E^2-\rvec p^2-(m-i\ga)^2&=-[\rvec p^2-(P+i\ga_E)^2]\cr
\then P+i\ga_E&=\sqrt{E^2-m^2+\ga^2+2im\ga}\cr
&=\sqrt{E^2-m^2}(1+{im\ga\over E^2-m^2}+\dots)\cr
\end{eq}Also mixed projections are possible, e.g.
systems with both nontrivial energy and momentum spread.

\section{Scalar Products for Hilbert  Spaces}

Unstable translation re\-pre\-sen\-ta\-tions involve 
positive type functions as an essential structure.
It is useful to illustrate the general structure
of the related scalar products   first on  finite dimensional
vector spaces.

\subsection{Dual Bases and Scalar Product Related Bases}

A vector space $V\cong\C^n$ has a basis $\{e^j\mid j=1,\dots,n\}$.
The dual vector space $V^T\cong\C^n$, i.e. the linear forms of $V$, has the dual
basis $\{\d e_j\mid j=1,\dots n\}$ leading to a decomposition of 
the identity transformation
\begin{eq}{l}
\hbox{dual product: }\dprod{\d e_k}{e^j}=\de^j_k,~~\id_V=e^j\ox\d e_j
\end{eq}With dual bases, a $V$-endomorphism $S:V\map V$
can be written as tensor product with its matrix components $S= S_j^k e^j\ox\d e_k$, 
$S^j_k\in\C$.
All this is purely algebraic - without any metrical structure.
There is no natural isomorphism between $V$ and its dual $V^T$.

A scalar product of $V$, sesquilinear and positive definite
\begin{eq}{l}
 d:V\x V\map \C,~~\left\{\begin{array}{l}
 d(v,w)=\ol { d(w,v)},~~ d(v,\al w)=\al  d(v,w),~\al\in\C\cr
v\ne0\iff d(v,v)>0\cr
 d(e^j,e^k)= d^{jk}\cr \end{array}\right.
\end{eq}endows $V$ with a Hilbert space structure.

 Since nondegenerate,
$\det  d^{jk}\ne 0$, it defines an antilinear isomorphism
between $V$ and its dual $V^T$.  Such a dual antilinear isomorphism is
the origin of Dirac's bra-ket notation 
\begin{eq}{l}
V\stackrel *\lrmap  V^T,~~e^j=\rstate{e^j}\lrmap \lstate{e^j}= d^{jk}\d e_k,~~
 d^{jk}=\sprod{e^j}{e^k}
 
\end{eq}With the scalar product induced   isomorphy
there is an antilinear isomorphy $\cong$ also  for
the tensor re\-pre\-sen\-ta\-tion of the endomorphisms, e.g. for the decomposition of
the identity which becomes antilinear isomorphic to the inverse scalar product
\begin{eq}{rcrcrl}
S&=& S_j^k e^j\ox\d e_k&\cong& S_j^k \rstate{e^j} d_{kl}\lstate{e^l}&=
S_{jl} \rstate{e^j}\lstate{e^l},~~S_{jl}=S_j^k d_{kl}\cr
\hbox{e.g. }\id_V&=&  e^j\ox\d e_j&\cong&  \rstate{e^j} d_{jl}\lstate{e^l}&
\end{eq}For a non-diagonal scalar product
the bra-ket decomposition of the identity is non-diagonal too.

The positivity of a scalar product can be expressed for
the characterizing matrix by the C*-algebra positivity
\begin{eq}{rl}
\hbox{complex $ (n\x n)$-matrix } d\succeq 0&\iff  d=\xi^*\o\xi\cr
&\iff  d= d^*\hbox{ and }\spec d \subnoteq \R_+
\end{eq}The hermitian $ d$ can be unitarily diagonalized
with positive diagonal elements (positive spectral values)
\begin{eq}{l}
 d=u^*\o\diag d\o u,~~u\in\SU(n)
\end{eq}For an  orthogonal basis  with individual 
normalization factors one obtains
\begin{eq}{l}
(\diag d)^{jk}=| d(j)|^2\de^{jk}  \hbox{ (no $j$-sum} )\then 
\id_V \cong\sum_j \rstate{e^j}{1\over | d(j)|^2}\lstate{e^j} 
\end{eq}By renormalization, an  orthonormal basis 
 can be used    
\begin{eq}{l}
 d=u^*\o\diag d\o u=\xi^*\o\bl1_n\o \xi \hbox{ with } \xi=\sqrt{\diag  d}\o u
\end{eq}

The scalar product defining  
positive type functions above, e.g. for unstable energy `eigenstates' 
\begin{eq}{l}
\sprod{\mu,\ga;E'}{\mu,\ga;E}
={1\over\pi}~{\ga\over (E-\mu)^2+\ga^2}\de({E-E'\over 2\pi}) 
\end{eq}are the distributional analogue to an orthogonal, but 
not normalized basis.
For particle collectives (below) - 
in a basis with the translation `eigenstates' -
 the scalar  product is even non-diagonal.

\subsection{Positive Type Functions}

Positive type functions are the continuous generalization of
scalar products for  finite dimensional spaces.

In contrast to compact groups, Hilbert re\-pre\-sen\-ta\-tions of a locally compact noncompact 
 group $G$  with Haar measure
have not to act on  square integrable functions.
The group algebra $L^1(G)$
contains the absolute integrable function classes on 
$G$. All function properties hold Haar measure almost everywhere.
A  Hilbert metric for $L^1(G)$
 uses  the group algebra dual, 
 the continuous linear forms $L^\infty(G)$
 with the essentially bounded function classes.
$L^\infty(G)$  has 
the  functions of positive type $d\in L^\infty(G)_+$ 
as a cone. Positive type
functions or, also, scalar product inducing functions
are defined by the property to give 
a positive linear form of the group algebra 
\begin{eq}{l}
 \hbox{for all }f\in L^1(G):~~\angle{\hat{f}*f}_d=
\int_{G\x G} dg_1dg_2~\ol{f(g_1)}~d(g^{-1}_1g_2)~f(g_2)
\ge 0\cr
\end{eq}Positive type functions have not to be positive functions.
A  continuous essentially bounded group function
$d\in L^\infty(G)$ is a positive type function 
if it gives rise to   finite  scalar product matrices
\begin{eq}{l}
d\in L^\infty(G)_+\iff
 d(g^{-1}_jg_k)_{j,k=1}^n \succeq 0
\hbox{ for all }n=1,2,\dots,~~g_{j,k}\in G\cr
\end{eq}It follows the conjugation invariance 
of $d$ and the absolute value restriction by the value at the neutral element  
\begin{eq}{l}
n=1:~~d(e)\ge 0\cr
n=2:~~(g_1,g_2)=(e,g),~{\scriptsize\pmatrix{
d(e)&d(g)\cr
d(g^{-1})&d(e)\cr}}\succeq 0\then\left\{\begin{array}{rl}
d(g^{-1})&=\ol{d (g)}\cr
|d(g)|&\le d(e)\end{array}\right.
\end{eq}

There is a  bijection between the positive type functions
and the cyclic Hilbert 
re\-pre\-sen\-ta\-tions\cite{GELRAI},  i.e. re\-pre\-sen\-ta\-tions 
with a cyclic vector $\rstate c$.
All positive type functions are re\-pre\-sen\-ta\-tion matrix elements
with a cyclic vector $d(g)=\sprod{c} {D(g)c}$.
 In general, the algebra $L^1(G)$ contains no unit, i.e.
 $\de_e\notin L^1(G)$ for the Dirac distribution. 
 The class of an approximate identity $\tilde \de_e$
leads to a cyclic vector whose re\-pre\-sen\-ta\-tion matrix element
is exactly the positive type function.

A state is a normalized positive  type function, i.e.
$d(e)=1$. There is a bijection 
between  the  pure states 
(extremal 
normalized positive type functions)
 and  the irreducible  Hilbert    re\-pre\-sen\-ta\-tions.
All pure states are irreducible re\-pre\-sen\-ta\-tion matrix elements
with a cyclic vector
(more exact formulation and more details in \cite{FOL,HELG2}).

In the former sections cyclic re\-pre\-sen\-ta\-tions of 
translations $\R^n$ have been used which are irreducible 
representations of 
affine subgroups $H\sx\R^n\sub\GL(\R^n)\sx\R^n$.
Positive type functions for cyclic
translation re\-pre\-sen\-ta\-tions  have the properties
\begin{eq}{l}
d\in L^\infty(\R^n)_+,~~
x\in\R^n\then  d(-x)=\ol{d(x)},~|d(x)|\le d(0)
\end{eq}The positivity of spectral values (diagonal elements) for 
finite matrices 
has its continuous counterpart - with Bochner's theorem \cite{BOCH} -
in the  positivity of  the Fourier transformed
positive type functions, i.e.
in positive spectral functions (positive Radon measures) for the
translation invariants (energy-momenta) $p\in\spec \R^n\cong\R^n$
\begin{eq}{l}   
d(x)=\int d^n p~\tilde d(p)e^{ipx}
\hbox{ with positive  distribution }\tilde d
\end{eq}With the energy-momentum distribution
the Hilbert space  can be defined via  energy-momentum functions
\begin{eq}{rl}
\sprod{f_1}{f_2}=\angle{\hat f_1*f_2}_d&=\int d^nx_1d^nx_2~\ol{f_1(x_1)}~d(x_2-x_1)~f_2(x_2) 
\cr&=\int d^np~\ol{\tilde f_1(p)}~\tilde d(p)~\tilde f_2(p) 
\end{eq}

The examples above for stable and unstable states (particles) are summarized in 
the following table
with positive   widths $\Ga,\ga>0$

{\scriptsize
\begin{eq}{c}

\begin{array}{|c|c|c|}\hline
\begin{array}{c}
\hbox{irreducible}\cr
\hbox{for group }\end{array}
&d\in L^\infty(\R^n)_+,~~x\mape d(x)&d(x)=\int d^n p~\tilde d(p) e^{ipx}\cr
\hline\hline
\R& t\mape e^{i\mu t}&\int dE~\de(E-\mu)e^{iEt}\cr\hline
\SO(3)\sx\R^3&\rvec x\mape {\sin Pr\over Pr}&\int {d^3 p\over 2\pi P}~\de(\rvec p^2-P^2) e^{-i\rvec p\rvec x}\cr\hline
\SO_0(1,3)\sx\R^4&
\begin{array}{r}
x\mape 2{\p\over\p m^2x^2}
\int d\psi[\vth(x^2)\cos|mx|\cosh\psi~\cr
+\vth(-x^2)e^{-|mx|\cosh\psi}]\end{array}
&\int {d^4 p\over 2\pi m^2}~\de( p^2-m^2)e^{ipx}\cr
\hline\hline
\O(1)\sx\R&t\mape e^{i\mu t-\Ga |t|}
&\int {d E\over \pi }~{\Ga\over (E-\mu)^2+\Ga^2} e^{iEt}\cr\hline
\O(3)\sx\R^3&\rvec x\mape {\sin Pr\over Pr}e^{-\ga r}
&\int {d^3 p\over \pi^2 }~{\ga\over (\rvec p^2-P^2+\ga^2)^2+4P^2\ga^2}
e^{-i\rvec p\rvec x}\cr\hline
\SO(1,3)\sx\R^4&x\mape
&\int {d^4 p\over \pi^2 |m| }~{\ga\over ( p^2-m^2+\ga^2)^2+4m^2\ga^2}
e^{i p x}\cr\hline

\end{array}
\end{eq}}

\noindent The Dirac energy-momentum distributions for stable structures lead to Hilbert
spaces with square integrable functions whereas the Breit-Wigner distributions
define the scalar products for the not square integrable functions used for
unstable structures.
The irreducible  translation re\-pre\-sen\-ta\-tion matrix elements 
$\R^n\ni x\mape e^{ipx}$ are pure states, i.e.
extremal 
normalized positive type functions.

\section{Probability Collectives of Unstable Particles}

Unstable particles come in collectives\cite{S02,BS03} as familiar from the
system with the two unstable neutral kaons
which - via nontrivial transition elements - `share one identity'. 
The scalar product can  involve nontrivial
(nondiagonal) transitions between unstable states (particles).

\subsection{Nonorthogonal Transition Probabilities}

A Hilbert space  operator $H$, e.g. a Hamiltonian,
 with eigenvectors
\begin{eq}{rll}
\lstate{E_2}H\rstate {E_1}&=E_1\sprod {E_2}{E_1}
&=\ol{E_2} \sprod {E_2}{E_1}\cr
\lstate{E_2}H^*\rstate {E_1}&=\ol{E_1}\sprod {E_2}{E_1}
&=E_2 \sprod {E_2}{E_1}\cr
\end{eq}has - if hermitian  - only real eigenvalues
and orthogonal eigenvectors for different eigenvalues
\begin{eq}{l}
H=H^*\then \left\{\begin{array}{l}
E=\ol E\cr
E_1\ne E_2\iff \sprod{E_2}{E_1}=0\end{array}\right.
\end{eq}If there arise  complex energy eigenvalues, i.e. $\spec H\not\subnoteq\R$,
 the Hamiltonian cannot be hermitian,
$H\ne H^*$.
Then, different eigenvectors have not to be orthogonal. 
Several eigenvectors may constitute a probability collective
with more than one dimension, 
i.e. a Hilbert subspace,
whose scalar product
is not decomposable with   eigenvectors.
The transition from a translation eigenstate basis $\{\rstate E\}$ to an orthogonal basis
$\{\rstate U\}$ 
defines a Hilbert-bein for the probability collective, a
non-unitary Hilbert space automorphism $\xi\notin\U(n)$, as 
a representative from the Hilbert-bein manifold $\GL(\C^n)/\U(n)$
\begin{eq}{l}
\rstate E=\xi \rstate U,~\sprod UU=\bl 1_n\then
\sprod EE=\xi^*\o\xi\cr
\end{eq}The scalar product matrix $\xi^*\o\xi$ is positive definite, but not
diagonal.

An example for an unstable particle collective is constituted by 
the  short and long  lived neutral kaon 
$\{\rstate E\}=\{K^0_{S,L}\}=\{\rstate S,\rstate L\}$
\begin{eq}{l}
m_S-m_L\sim 35\x 10^{-13}{\rm MeV\over c^2},~~\left\{\begin{array}{lcr}
\Ga_S&\sim &72\x 10^{-13}{\rm MeV\over c^2}\cr
\Ga_L&\sim &0.13\x 10^{-13}{\rm MeV\over c^2}\cr\end{array}\right.
\end{eq}which are  related to the 
orthonormal CP- or time reflection eigenstates 
$\{\rstate U\}=\{K^0_\pm\}=\{\rstate +,\rstate -\}$.
 A  $2\x 2$-Hilbert-bein
$\xi\in\GL(\C^2)/\U(2)$  can be parametrized with two complex numbers
\begin{eq}{l} 
{\scriptsize\pmatrix{ \rstate S\cr \rstate L\cr}}
=\xi{\scriptsize\pmatrix{ \rstate +\cr \rstate -\cr}},~~
\xi
=
{1\over N\sqrt{1+|\ep|^2}}
{\scriptsize\pmatrix{1&\ep\cr \ep&1\cr}}
,~~\ep,N\in\C\cr
\xi^*\o\xi={\scriptsize\pmatrix{\sprod SS&\sprod SL\cr \sprod LS&\sprod LL\cr}}=
{1\over |N|^2}{\scriptsize\pmatrix{1&\de\cr \de&1\cr}},~~\spec \xi^*\o \xi=\{
{1\pm \de\over |N|^2}\}
\cr
\de={\ep+|\ep|\over 1+|\ep|^2}\sim 0.327\x10^{-2}\hbox{ (experimental value)}
\then \xi \notin\U(2)\cr

\end{eq}

For an understanding of the actual re\-pre\-sen\-ta\-tions used
and their Schur-non\-or\-tho\-go\-na\-li\-ty 
the re\-pre\-sen\-ta\-tions of the affine group 
$\O(1)\sx\R$ as embedded in a relativistic
spacetime structure has to be considered.
That will not be done here. The related 
invariants (masses and widths)  and the 
nonorthogonal scalar product matrices are 
assumed as experimentally given. 

\subsection{Collectives of Unstable Energy States}

The time translation matrix elements for a collective with 
$n$ unstable states 
come in $(n\x n)$-matrices
\begin{eq}{rl}
\R\ni t\mape&  
=\xi^*\o \diag
\int {dE\over \pi}{\Ga_j\over (E-\mu_j)^2+\Ga_j^2}
e^{iEt}\o \xi\cr
&=\xi^*\o \diag
e^{i\mu_jt-\Ga_j|t|}\o \xi\cr
\diag {1\over\pi}{\Ga_j\over (E-\mu_j)^2+\Ga_j^2}
&={1\over\pi}\scriptsize{\pmatrix{
{\Ga_1\over (E-\mu_1)^2+\Ga_1^2}
&0&\dots&0\cr
0&{\Ga_2\over (E-\mu_2)^2+\Ga_2^2}&\dots&0\cr
&\dots&\dots&\cr
0&\dots&0&{\Ga_n\over (E-\mu_n)^2+\Ga_n^2}\cr}}\cr

\end{eq}A Hilbert-bein $\xi\notin\U(n)$  relates to each other
 an energy-eigenstate basis and  an orthogonal one.
 The time translation matrix elements for the neutral translation $t=0$ 
give the scalar product matrix $\xi^*\xi$.
 
E.g., the explicit matrix  for the kaon collective
with  invariants $m_{S,L}\pm i\Ga_{S,L}$
looks as follows
\begin{eq}{r}
{1\over |N|^2(1+|\ep|^2)}
\int{dE\over\pi}
{\left(\begin{array}{c|c}
{\Ga_S\over (E-m_S)^2+\Ga_S^2}
+|\ep|^2{\Ga_L\over (E-m_L)^2+\Ga_L^2}&
\ep{\Ga_S\over (E-m_S)^2+\Ga_S^2}
+\ol \ep{\Ga_L\over (E-m_L)^2+\Ga_L^2}\cr\hline
\ol \ep{\Ga_S\over (E-m_S)^2+\Ga_S^2}
+ \ep{\Ga_L\over (E-m_L)^2+\Ga_L^2}&
{\Ga_L\over (E-m_L)^2+\Ga_L^2}
+|\ep|^2{\Ga_S\over (E-m_S)^2+\Ga_S^2}\cr\end{array}\right)}e^{iEt}\cr\cr
={1\over |N|^2(1+|\ep|^2)}
{\left(\begin{array}{c|c}
e^{im_St-\Ga_S|t|}+|\ep|^2e^{im_Lt-\Ga_L|t|}&
\ep e^{im_St-\Ga_S|t|}+\ol\ep e^{im_Lt-\Ga_L|t|}\cr\hline
\ol \ep e^{im_St-\Ga_S|t|}+\ep e^{im_Lt-\Ga_L|t|}&
e^{im_Lt-\Ga_L|t|}+|\ep|^2e^{im_St-\Ga_S|t|}\cr\end{array}\right)}
\end{eq}

As in general for unstable `states', 
a distributive basis needs all energies for all 'eigenstates'
\begin{eq}{l}
\Ga_j>0:~~\{\rstate{\mu_j,\Ga_j;E}\mid E\in \R, j=1,\dots,n\}\cr
\end{eq}The scalar product distribution is the positive re\-pre\-sen\-ta\-tion matrix function
\begin{eq}{l} 
\sprod{\mu_l,\Ga_l;E'}{\mu_k,\Ga_k;E}
=\xi^*\o{1\over\pi} \diag {\Ga_j\over (E-\mu_j)^2+\Ga_j^2}\o \xi~~
\de({E-E'\over 2\pi}) 
\end{eq}The Hilbert space vectors uses functions for
all 'eigenstates'
\begin{eq}{l}
\rstate{\mu,\Ga;f}={\SUM_{j=1}^n}\int dE~ f_j(E) \rstate{\mu_j,\Ga_j;E}
\end{eq}

\subsection{Collectives  of Unstable Particles}

The Lorentz group compatible extension for scalar particle collectives 
\begin{eq}{rl}
\R^4\ni x\mape&  
\xi^*\o \diag
\int {d^4p\over 2\pi}{1\over m_j^2}\de_{\ga_j}(p^2-m_j^2)
e^{ipx}\o \xi\cr
\diag {1\over m_j^2}\de_{\ga_j}(p^2-m_j^2)&={1\over \pi}\scriptsize{\pmatrix{
{{2\ga_1\over m_1}\over(p^2-m_1^2+\ga_1^2)^2
+4m_1^2\ga_1^2}&\dots&0\cr
\dots&\dots&\dots&\cr
\dots&\dots&\dots&\cr
0&\dots&{{2\ga_n\over m_n}\over(p^2-m_n^2+\ga_n^2)^2
+4m_n^2\ga_n^2}\cr}}
\end{eq}leads to a Hilbert space with a distributive basis
with  'eigenstates' of mass $m_j^2$ and widths $\ga_j$
 for all energy-momenta 
\begin{eq}{l}
\ga_j>0:~~\{\rstate{m^2_j,\ga_j;p}\mid p\in \R^4, j=1,\dots,n\}\cr
\end{eq}The scalar product distribution
involves the positive re\-pre\-sen\-ta\-tion matrix function
\begin{eq}{l} 
\sprod{m^2_l,\ga_l;p_2}{m^2_k,\ga_k;p_1}
=\xi^*\o {1\over\pi}\diag 
{{2\ga_j\over  m_j}\over(p^2-m_j^2+\ga_j^2)^2
+4m_j^2\ga_j^2}\o\xi~~2\pi \de(p_1-p_2) \cr
\end{eq}The Hilbert space vectors uses functions for
all 'eigenstates' in a nondecomposable collective 
\begin{eq}{l}
\rstate{m^2,\ga;f}={\SUM_{j=1}^n}\int {d^4p\over 2\pi}
 f_j(p) \rstate{m^2_j,\ga_j;p}
\end{eq}

\noindent{\bf Acknowledgments} 
\vskip1mm
\noindent I gratefully acknowledge theoretical and
experimental  discussions with
Florian Beissner and Walter Blum.


\begin{thebibliography}{99}

\bibitem{BOCH}{S. Bochner, Monotone Funktionen, Stieltjessche Integrale und
harmo\-ni\-sche Analyse, {\it Math. Ann.}, 108 (1933), 378-410}




\bibitem{LIE13}{N. Bourbaki, {\it Lie Groups and Lie Algebras, Chapters 1-3} (1989),
Springer, Berlin, Heidelberg, New York, London, Paris, Tokyo}
 



\bibitem{FOL}{G.B. Folland, {\it A Course in Abstract Harmonic Analysis}
(1995), CRC Press, Boca Raton, Ann Arbor, London, Tokyo} 

\bibitem{GELRAI}{I.M. Gelfand and D.A. Raikov, 
Irreducible unitary re\-pre\-sen\-ta\-tions of locally bicompact groups, 
{\it Mat. Sb.} 13 (55) (1942), 301-316}


\bibitem{GEL1}{I.M. Gel'fand, G.E. Shilov,
{\it Generalized Functions I (Properties and Operations)}
(1958, English translation 1963), Academic Press, New York and London}

\bibitem{HELG2}{S. Helgason, {\it Groups and
       Geometric Analysis} (1984), Academic Press, 
        New York. London, Sydney, Tokyo, Toronto, etc.}

\bibitem{KIR}{A.A. Kirillov, {\it
Elements of the Theory of Re\-pre\-sen\-ta\-tions} (1976), Springer-Verlag, Berlin,
Heidelberg, New York}


\bibitem{KNAPP}{A. Knapp, {\it Representation Theory of Semisimple Groups}
(1986), Princeton University Press, Princeton}



\bibitem{MACK1}{G.W. Mackey, On induced re\-pre\-sen\-ta\-tions of groups
 (1951), Amer. J. Math. 73, 576-592}

\bibitem{MACKP}{G.W. Mackey, {\it Induced Re\-pre\-sen\-ta\-tions of Group and Quantum
Mechanics} (1968), W.A. Benjamin, New York, Amsterdam}

\bibitem{NEUM}{J. von Neumann, {\it Collected Works} (1961), Pergamon, New York}



\bibitem{PWEYL}{F. Peter, H. Weyl, 
Die Vollst\"andigkeit der primitiven Darstellungen einer geschlossenen
kontinuierlichen Gruppe,
{\it Math. Ann.} 97 (1927), 737-755}


\bibitem{SCHUR}{I. Schur, Neue Begr\"undung der Gruppencharaktere,
{\it Sitzungsber. Preuss. Akad. 1905}, 406} 


\bibitem{VIL}{N.Ja.  Vilenkin, A.U. Klimyk,
{\it Re\-pre\-sen\-ta\-tions of Lie Groups and Special Functions} (1991),
Kluwer Academic Publishers, Dordrecht, Boston, London}



\bibitem{WIG}{E. P. Wigner, 
On unitary re\-pre\-sen\-ta\-tions of the inhomogeneous Lorentz group,
 {\it Annals of Mathematics} 40 (1939), 149-204}

\bibitem{S96}{H. Saller,
Analysis of Time-Space Translations in Quantum Fields,
{\it International Journal of Theoretical Physics} {\bf 36}, 1033-1071 (1997)}

\bibitem{S012}{H. Saller,
 On the Relation between Time Re\-pre\-sen\-ta\-tions and Inner Product Spaces
 {\it International Journal of Theoretical Physics},
 published, {hep-th/0112208 }}

\bibitem{S02}{H. Saller,
 Probability Collectives for Unstable Particles
{\it International Journal of Theoretical Physics}
{\bf 40},1491-1509 (2002) {hep-th/0205059 }}


\bibitem{BS03}{}{W. Blum and H. Saller,
Relativistic resonances as non-orthogonal states in Hilbert-space.
{\it Eur. Phys. J.C.} {\bf 28}, 279-295 (2003)}


\bibitem{S041}{H. Saller,
 The basic physical Lie operations, hep-th/0410147} 






 \end{thebibliography}
\end{document}